\documentclass{PNAStwo}

\usepackage{graphicx}
\usepackage{amssymb,amsfonts,amsmath}
\usepackage{bm}

\begin{document}

%%%
\conflictofinterest{Conflict of interest footnote placeholder}

\track{Insert 'This paper was submitted directly to the PNAS office.' when applicable.}
%%%

\title{Structural Change of Myosin Motor Domain and Nucleotide Dissociation}

\author{Fumiko Takagi\thanks{To whom correspondence should be addressed. E-mail: fumiko@cp.cmc.osaka-u.ac.jp}\affil{1}{Formation of Soft Nanomachines, Core Research for Evolutional Science and Technology,  Japan Science and Technology Agency, Yamadaoka, Suita, Osaka 565-0871, Japan}%
\affil{2}{Cybermedia Center Osaka University, Toyonaka, Osaka 560-0043, Japan}, \and
Macoto Kikuchi\affil{1}{}\affil{2}{}
}
% 

%\date{\today}

\maketitle

\begin{article}

\begin{abstract}
We investigated the structural relaxation of myosin motor domain 
from the pre-power stroke state to the near-rigor state
using molecular dynamics simulation of a coarse-grained protein model.
To describe the structural change, we propose a ``dual G\={o}-model,'' 
a variant of the G\={o}-like model that has two reference structures.
The nucleotide dissociation process is also studied by introducing 
a coarse-grained nucleotide in the simulation.
We found that the myosin structural relaxation toward the near-rigor conformation
cannot be completed before the nucleotide dissociation.
Moreover, the relaxation and the dissociation occurred cooperatively
when the nucleotide was tightly bound to the myosin head.
The result suggested that the primary role of the nucleotide is
 to suppress the structural relaxation.
\end{abstract}

%%%
%%%

\section{Introduction}

The mechanism of biomolecular motors is one of the major topics in biophysics.
Among a number of such systems that have been found so far, 
the actomyosin motor is of a particular interest, 
because it is responsible for muscle contraction 
and cellular movements in eukaryotic cells.
Myosin moves unidirectionally along the actin filament 
using chemical energy released by ATP hydrolysis
\cite{AnnuRevBiochem-68_687, BIOPHYS-1_1, NatRevMolCellBiol-2_387}.
It is widely recognized that the efficiency of this energy conversion 
is very high compared with macroscopic artificial machines, 
in spite of the fact that biomolecular motors work 
under a noisy environment in the cell.
In fact, the free energy released during each ATP hydrolysis is only about $20k_BT$;
therefore the thermal fluctuation should be appreciable.
Although recent progress in imaging 
and of nanomanipulation has enabled the observation of single molecules,
the movement mechanism of the actomyosin motor is still not understood.

There has been a long-standing controversy 
between the tight-coupling (lever-arm) model and the loose-coupling model.
X-ray crystallographic studies have revealed that the angle of the neck domain 
changes relative to the motor domain, depending on the nucleotide state.
The ``lever-arm'' model was proposed based on these observations, 
in which the structural change of myosin is tightly coupled 
with the ATP hydrolysis cycle, 
and directly causing a stepwise sliding motion.
It was shown, however, 
that the sliding distance of the myosin along the actin filament per ATP 
at the muscle contraction can be much longer than the displacement 
predicted by the lever-arm-like structural change 
of a single myosin molecule \cite{Nature-316_366}.
Moreover, it is questionable 
whether a material as soft as proteins can accurately switch its conformation 
in the same way as a macroscopic machine under thermal fluctuation.

In the ``loose-coupling'' model, in contrast, the structural change 
does not always correspond to a step in a one-to-one correspondence;
the motion is intrinsically stochastic and thermal fluctuation 
is an essential ingredient for its mechanism \cite{AdvBiophys-26_97}.
The simplest class of models that produce the loose-coupling mechanism 
is based on a thermal ratchet, 
in which a myosin molecule is treated as a Brownian particle 
that moves along a periodic and asymmetric potential 
under both thermal noise and non-thermal perturbations
\cite{RevModPhys-69_1269,PhysRep-361_57}.
Although ratchet systems can, in fact, 
exhibit unidirectional flow in a noisy environment,
a high efficiency comparable to that of the actomyosin system 
is found difficult to achieve using only a simple ratchet system.
Even if the ratchet models do to express some essence 
of the mechanism of the biomolecular motor, they are too much simplified 
and we should say that the connection with the real actomyosin system 
is rather vague.
In particular, since the myosin is expressed as a particle, 
the effect of its conformational change is, 
at best, only implicitly taken into account.
A somewhat more realistic modelling is desirable,
which can reflect the chain conformation. 

Recently, it was revealed by single molecule experiments 
that the chemomechanical cycle of the myosin head is controlled by a load 
on the actomyosin crossbridge \cite{NCB-5_980,NCB-7_861,Cell-116_737}.
The observation suggested that the rate of ADP release from the myosin head 
depends on the force acting on myosin; namely, 
the chemical reaction rate varies with the deformation of the myosin.
If the myosin head indeed acts as such a ``strain sensor,''
this would be reminiscent of a classical model by A. Huxley
\cite{ProgBiophysBiophsicalChem-7_255};
in this model, the myosin head is supposed to undergo Brownian motion 
and change into a tightly bound state to the actin filament 
triggered by a structure dependent chemical reaction.

The relationship between structure and function of proteins
has long been investigated.
Thus far, mainly the static aspects of proteins have been considered,
for example, the classical ``lock and key'' model of an enzyme.
Recently, the role of structural fluctuations,
or that of a more drastic structural change, including ``partial unfolding,''
on protein functions has become a subject of growing interest.
Although there have been many experimental studies to clarify 
the dynamical processes of protein at a functional level, it is still difficult 
to observe the structural changes with high resolution in both space and time.
Computer simulations serve as possible alternatives.
Typical computational studies treated equilibrium fluctuations 
near a crystal structure using the all-atom model \cite{JMB-340_345,PRL-94_078102} 
or the elastic network model \cite{PNAS-100_13253};
although these types of simulations cannot deal with large-scale structural change,
the low-frequency fluctuation modes were found 
to be consistent with the direction of motion of the structural change 
associated with the function.

Some attempts have been made to simulate a larger structural change 
beyond the elastic regime using a class of models called the G\={o}-like model
\cite{PNAS-103_5367}.
According to the recently developed theory of spontaneous protein folding,
the protein energy landscape has a funnel-like global shape 
toward the native structure.
The G\={o}-like model is certainly the simplest class of models 
that can realize a funnel-like landscape
\cite{AnnuRevBiophysBioeng-12_183,AnnuRevPhysChem-48_545} 
and has successfully described the folding process of small proteins.
It is, however, not suitable for the study of a change between two conformations, 
because only the interaction between the pairs of residues 
that contact each other in the native state are taken into account;
the conformation other than the native state becomes too unstable as a result.
Thus, a model is desirable in which two conformations can be embedded.
Here, we introduce a new model bearing this property 
as a variant of the G\={o}-like model.

In this paper, 
we investigate the dynamics of the myosin conformational change 
from the pre-power stroke state to the near-rigor state 
by molecular dynamics simulations of a coarse-grained protein.
This process is called ``power stroke''
because the angle of the lever-arm changes remarkably,
and it is considered in the lever-arm model 
that this structural change directly causes the force generation.
To describe the structural change, we propose the "dual G\={o}-model."
The dissociation process of the nucleotide that accompanies 
the conformational change is also involved in the simulation 
by introducing a coarse-grained nucleotide.
To our knowledge, 
the ligand at the binding site has not been considered explicitly 
in coarse-grained protein simulations,
possibly because the primary role of ATP is considered to be the release 
of chemical energy through hydrolysis,
and the excluded volume effect of the molecule has not been investigated.
We, however, consider that the presence or absence of nucleotides 
in the binding site would profoundly affect the structural fluctuation of the protein.
Therefore, it is important to perform simulations including the nucleotide.

\section{Results}

First, we introduce the dual G\={o}-model.
While only the native structure is taken as a reference structure 
for the potential energy function in the standard G\={o}-like models, 
the dual G\={o}-like model takes two reference structures, structure 1 and structure 2, 
in the effective potential.
For the interaction between ``native-contact'' pairs,
each potential energy function has two minima corresponding to two reference structures.
To study the relaxation process from structure 2 to structure 1, 
the minimum corresponding to structure 2 is given a slightly higher energy than that of structure 1 to make structure 1 more stable.
In this study,
we used a model based on one of the C$_{\alpha}$ G\={o}-like models
\cite{JMB-298_937, JMB-313_171}, which involves local interactions such as bond length, bond angle, and dihedral angle interactions as well as the native-contact interaction.
For these local interactions, 
we set that each potential energy function has two minima of the same depth
corresponding to two reference structures.

\begin{figure}
\begin{center}
% \begin{tabular}{ll}
%   (a) structure-1: near-rigor
%   &(b) structure-2: pre-powerstroke\\
% \resizebox{4cm}{!}{\includegraphics{1q5g.eps}}
% &\resizebox{4cm}{!}{\includegraphics{1vom.eps}}\\
% \end{tabular} 
\includegraphics{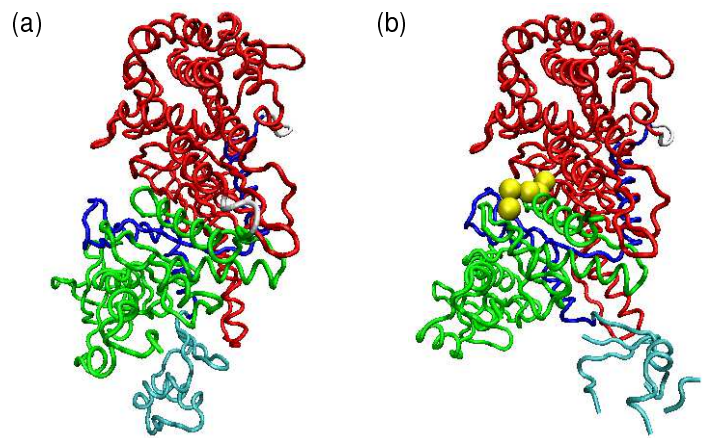}
 \caption{We chose (a) near-rigor structure, 1Q5G, 
 and (b) pre-power stroke structure, 1VOM
 for structure 1 and 2, respectively.
 1VOM contains the ADP$\cdot$P$_{\mbox{\small i}}$ analog, ADP$\cdot$VO$_4$ (yellow).
 Also shown are the N-terminal (green), 50 kDa subdomain (red)
 and the converter (cyan) included in C-terminal subdomain (blue)
 that is connected to the lever arm.
 }\label{Fig:X-ray_structures}
\end{center} 
\end{figure}

As reference structures,
we choose X-ray crystallography structures of {\it Dictyostelium discoideum} myosin II:
the near-rigor structure without nucleotide, 1Q5G \cite{NSB-10_826} for structure 1 
and the pre-power stroke structure with ADP$\cdot$P$_{\mbox{\small i}}$ analog,
 1VOM \cite{Biochem-35_5404} for structure 2 (Fig. \ref{Fig:X-ray_structures}).
The initial structure is the pre-power stroke structure with a coarse-grained nucleotide (Fig. \ref{Fig:ATP}) located at the nucleotide-binding site.

\begin{figure}
\begin{center}
 \includegraphics{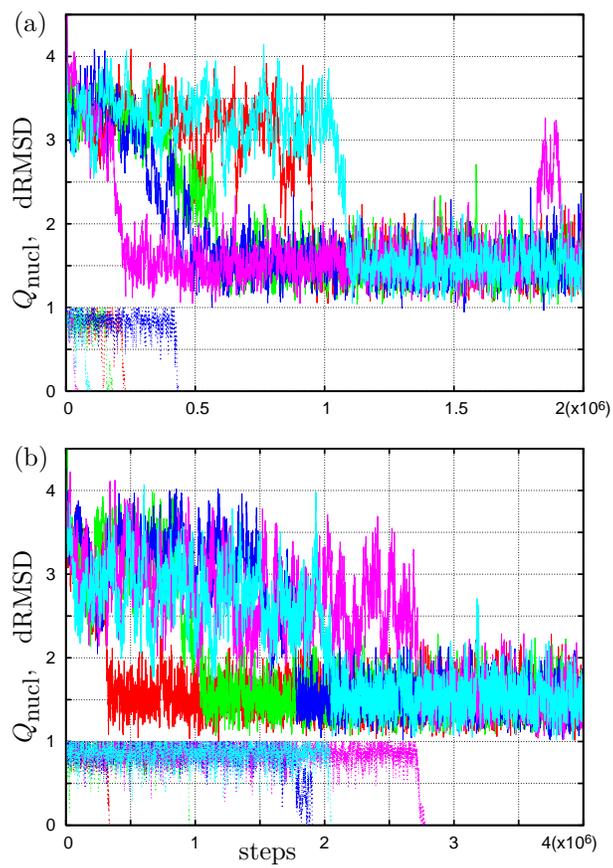}
 \caption{Relaxation time courses of dRMSD (from 1Q5G) (solid line) and
 $Q_{\mbox{\small nucl.}}$ (dotted line) for five trajectories are shown.
 Different colors distinguish different runs. 
 (a) $k_{\mbox{\small p-n}}=0.6$, (b) $k_{\mbox{\small p-n}}=0.7$}
 \label{Fig:time_courses}
\end{center}
\end{figure}

Typical time courses of the distance root mean square deviation (dRMSD) are shown in Fig. \ref{Fig:time_courses}
for (a) $k_{\mbox{\small p-n}}=0.6$ and (b) $0.7$,
where $k_{\mbox{\small p-n}}$ is the strength parameter of the protein-nucleotide interaction.
The dRMSD from the near-rigor structure is defined as
\begin{equation}
 {\mbox{dRMSD}}= \sqrt{\frac{2}{N (N-1)}\sum_{i<j} (r_{ij}-r_{ij}^{(1)})^2},
\end{equation}
in which $r_{ij}=|{\mathbf r}_{ij}|=|{\mathbf r}_i\!-\!{\mathbf r}_j|$ is the distance between C$_{\alpha}$ carbons of the $i$th and $j$th residues in the given conformation, 
and ${r_{ij}^{(1)}}$ indicates their distance in the near-rigor structure.
At the initial conformation, the dRMSD of $\sim$ 3.9\AA,
and decreases rapidly to $\sim$ 3\AA,
and stays there for a while.
The conformation eventually relaxes into the final state, dRMSD $\sim$ 1.5\AA.
This final state is actually the near-rigor state,
judging from its average structure;
in fact, the dRMSD of the average structure is $\sim$ 1\AA.
In short, the myosin motor domain in the pre-power stroke conformation relaxes at first into the intermediate state and then to the near-rigor conformation.

\begin{figure}
 \begin{center}
  \includegraphics{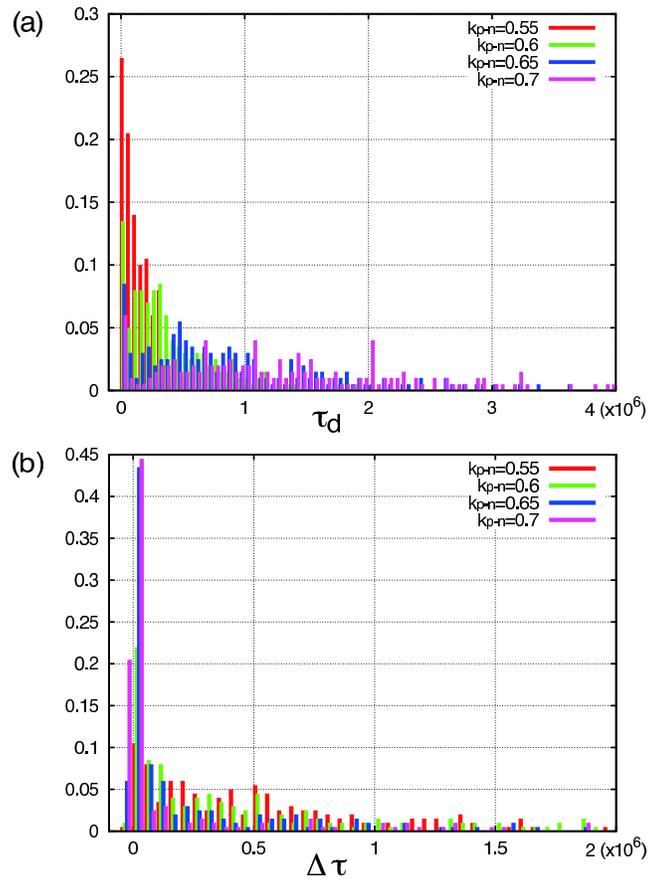}
\caption{Histogram of (a) the number of steps before dissociation and (b) the delay of the relaxation after the dissociation 
from 200 independent runs for each $k_{\mbox{\small p-n}}$.}
\label{Fig:hist_diff}
 \end{center}
\end{figure}

We introduced an index to characterize the state of the nucleotide binding,
$Q_{\mbox{\small nucl}}(\Gamma)$;
we count how many of the nucleotide contacts that is formed in structure 2 
(the pre-power stroke) remain in a given conformation, $\Gamma$.
Then, $Q_{\mbox{\small nucl}}(\Gamma)$ is defined as this number divided by the number of nucleotide contacts in structure 2.
$Q_{\mbox{\small nucl}}\sim$ 1 when a nucleotide is bound,
and $Q_{\mbox{\small nucl}}=0$ if the nucleotide-binding site is empty.
The time courses of $Q_{\mbox{\small nucl}}$ are also shown in Fig. \ref{Fig:time_courses}.
The structural relaxation occurs after or, at the earliest,
at the same time as the nucleotide dissociation.
Furthermore, the relaxation tends to synchronize with the dissociation as $k_{\mbox{\small p-n}}$ increases.
To clarify the $k_{\mbox{\small p-n}}$ dependency of the synchronization,
we plotted the histograms of $\tau_{\mbox{\small d}}$ 
and of $\Delta \tau =\tau_{\mbox{\small d}}-\tau_{\mbox{\small r}}$
from 200 independent runs for each value of $k_{\mbox{\small p-n}}$,
where $\tau_{\mbox{\small d}}$ is the number of steps taken before the nucleotide dissociates,
$\tau_{\mbox{\small r}}$ is the number of steps taken 
before the conformation relaxes to the near-rigor state,
and $\Delta \tau$ is the delay in the relaxation after the dissociation takes place
(Fig. \ref{Fig:hist_diff}).

For small $k_{\mbox{\small p-n}}$, 
both the histograms of both $\tau_{\mbox{\small d}}$ 
and of $\Delta \tau$ show exponential decay;
thus, the nucleotide dissociation and the relaxation of the myosin conformation 
are considered to be decoupled.
For larger $k_{\mbox{\small p-n}}$, on the other hand,
the histogram of $\tau_{\mbox{\small d}}$ cannot be fitted to an exponential decay.
In addition, the average of $\tau_{\mbox{\small d}}$ is shifted to the right and the delays, $\Delta \tau$ become shorter;
in other words, the nucleotide is unbound later and the conformational relaxation tends to occur immediately after the nucleotide dissociation.
For $k_{\mbox{\small p-n}}=0.7$,
dissociation and relaxation occur nearly simultaneously in over 60\% of 200 trajectories.
Note that apparent ``$\Delta \tau < 0$ cases'' are caused simply
from the numerical ambiguity of $\tau_{\mbox{\small d}}$ and $\tau_{\mbox{\small r}}$  
and actually correspond to coincidental dissociation-relaxation.

\begin{figure}
 \begin{center}
  \includegraphics{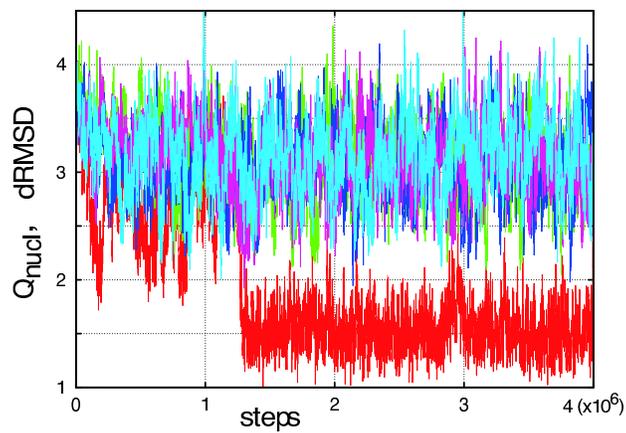}
 \caption{Time sequence of dRMSD. The red line is the trajectory of the no-constraint simulation, and the other lines are trajectories of simulations in which nucleotide-binding site are constrained.}
 \label{Fig:nucl_constraint}
 \end{center}
\end{figure}

The largest difference between the intermediate state 
and the initial (pre-power stroke) conformation is 
the position of the converter domain with respect to the other subdomains;
the relative positions among other subdomains 
(for example, the N terminal and the 50-kDa subdomain)
are similar to those in the pre-power stroke conformation.
The average dRMSD of the intermediate state varies slightly 
with the parameter $k_{\mbox{\small p-n}}$ (Fig. \ref{Fig:nucl_constraint}), 
reflecting little difference in the position of the converter relative to the other subdomains.

\begin{figure}
\begin{center}
  \resizebox{5cm}{!}{\includegraphics{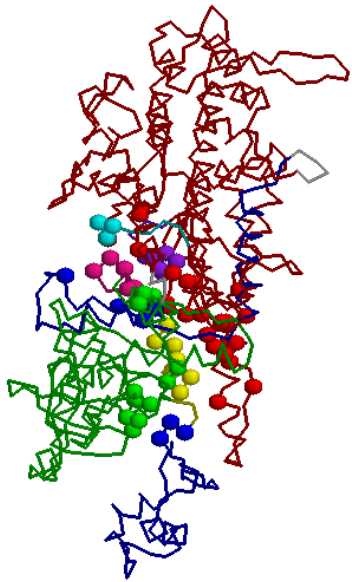}}
 \caption{Residues included in the contacts that are formed at the final relaxation
to the near-rigor.}
 \label{Fig:contact_drmsd_corr}
\end{center}
\end{figure}

Some of the native contacts of structure 1 are not formed until the conformation finally relaxes to the near-rigor state.
Figure \ref{Fig:contact_drmsd_corr} shows the residues included in these contacts.
They are concentrated at the boundary between the N terminal and the 50-kDa subdomain,
that is, the region around the nucleotide-binding site.
Thus, the final relaxation process consists of a rearrangement 
of the N terminal subdomain against the other part.

As already mentioned,
the myosin motor domain relaxes to the near-rigor conformation 
only after the dissociation of the nucleotide and not before.
Thus, it seems that the nucleotide must be unbound for the final relaxation to occur.
This observation leads to a speculation that the nucleotide blocks the deformation 
of myosin around the nucleotide-binding site by its volume.
To investigate the case of when the nucleotide cannot dissociate,
we attempted a nucleotide-free and constrained simulation.
In this simulation,
instead of treating the nucleotide molecule explicitly,
we connected the residues that would contact with the nucleotide by virtual bonds,
to force the nucleotide-binding site to keep the pre-power stroke form.
Figure \ref{Fig:nucl_constraint} shows the time courses obtained by the simulations.
We find that the conformation remains at the intermediate state 
and that the relaxation toward the near-rigor state is not completed.

\section{Discussion}

The relaxation simulation using coarse-grained myosin and the nucleotide
have shown that the myosin motor domain does not relax to the near-rigor conformation 
before the nucleotide dissociates.
Ishijima {\it et al.} \cite{Cell-92_161} showed 
by simultaneous observation of ADP release and mechanical events that 
force is generated at the same time as or 
several hundreds of milliseconds after the dissociation of ADP.
Our results are consistent with their experimental findings
if force generation is preceded by structural relaxation.
Moreover, the results from the simulations 
in which the conformation of the nucleotide-binding site is constrained 
also indicate that the relaxation is indeed prevented if the nucleotide cannot dissociate.

Based on these observations, we now suggest %a possibility 
that the primary role of the nucleotide in the ``power stroke'' process
is to suppress relaxation through blocking deformation around the nucleotide-binding site by its volume.
In this scenario, hydrolysis is required to alter the affinity of the nucleotide to the binding site. 
In particular, our simulations have shown that
the structural relaxation is synchronous with nucleotide dissociation 
when the nucleotide is tightly bound to the myosin head.
In other words, the nucleotide dissociates cooperatively with
the motion of the subdomain.
This strong coupling of deformation and dissociation
seems to be relevant to the function of the ``strain sensor,'' 
in which nucleotide dissociation 
is controlled by the strain induced by an external force.
The correlation depends on the binding strength, $k_{\mbox{\small p-n}}$;
the relaxation is only loosely coupled with the dissociation 
for weak binding conditions.
The origin of a large kinetic diversity among myosins 
\cite{CurrOpinCellBiol-16_61, %% review:"Relating biochemistry and function in the myosin superfamily"
BBAMolCellRes-1496_3} %% review: "Myosins: a diverse superfamily"
may be attributed to this binding strength dependence.

The intermediate state observed in the relaxation process should also be discussed.
Although several intermediate states have been revealed by 
kinetic experiments\cite{CurrOpinCellBiol-16_61},
their structural aspects,
except for ADP$\cdot$P$_{\mbox{\small i}}$ state, are little known.
Shih {\it et al}. \cite{Cell-102_683} reported, from their FRET study, that there are two ``pre-power stroke'' conformations;
while one conformation corresponds to the crystal structure 
of the complex with the ADP$\cdot$P$_{\mbox{\small i}}$ analog, 
the other conformation has not yet been observed using crystallography.
We found that the average structure of the intermediate state observed in the present study is consistent with the latter conformation.

%The constrained simulation using virtual bonds instead of the coarse-grained nucleotide could revealed a role of the nucleotide on the myosin structural change.
%Similar approach will also be effective in understanding the actin activation of ATPase or other effects of ligands.
%
Our dual G\={o} model 
is effective in studying myosin conformational changes.
Recently, a similar approach was proposed 
for a Monte Carlo simulation of a lattice protein,
in which a double-square-well potential for native-contact pairs was introduced%
\cite{JPhysChemB-108_5127}.
This type of model, 
in which two conformations are embedded in an energy potential function,
will be useful in understanding the dynamics of protein conformational change.

\section{Models and Methods}

Our dual G\={o}-model
is a variant of the C$_{\alpha}$ G\={o}-like model \cite{JMB-298_937,JMB-313_171}.
A protein chain is formed, consisting of spherical beads 
that represent C$_{\alpha}$ atoms of amino acid residues connected by virtual bonds.
In conventional G\={o}-like models, 
only amino acid pairs that contact in the native conformation are assigned 
an effective energy.
In the dual G\={o}-model, on the other hand,
the effective energy function takes two reference structures,
structure 1 and structure 2.
A nucleotide molecule is also expressed as a chain of connected beads.

The total energy of the system, $U_{\mbox{\small tot}}$ consists of three terms;
\begin{equation}
 U_{\mbox{\small tot}} = U_{\mbox{\small p}} + U_{\mbox{\small n}} + U_{\mbox{\small p-n}},
  \label{EQ:potential_total}
\end{equation}
where $U_{\mbox{\small p}}$ is the intraprotein interaction, 
$U_{\mbox{\small n}}$ is the intranucleotide interaction,
and $U_{\mbox{\small p-n}}$ is the interaction between protein and nucleotide.

The effective protein energy $U_{\mbox{\small p}}$
at a conformation $\Gamma$ is given as,
\begin{equation}
    U_{\mbox{\small p}}(\Gamma,\Gamma^{(1)},\Gamma^{(2)})
     = U^{\mbox{\small b}}
     + U^{\theta}
     + U^{\phi}
     + U^{\mbox{\small nc}}
     + U^{\mbox{\small nnc}}.
     \label{EQ:protein_potential}
\end{equation}
where $\Gamma^{(1)}$ and $\Gamma^{(2)}$ stand for the conformations of the two reference structures.
The terms in Eq. \ref{EQ:protein_potential} are defined as follows:
\begin{equation}
  U^{\mbox{\small z}} = \sum_{i}
   \min\{V_{\mbox{\small z}}^{(1)}(z_{i}),
   V_{\mbox{\small z}}^{(2)}(z_{i})\},
\end{equation}
\begin{equation}
  U^{\mbox{\small nc}} = \sum^{\small\begin{minipage}{3.2em}\setlength{\baselineskip}{0mm} native\\contact\end{minipage}}_{j<i-3} 
  \min\{V_{\mbox{\small nc}}^{(1)}({\mathbf r}_{ij}),
   C_{12} V_{\mbox{\small nc}}^{(2)}({\mathbf r}_{ij})\},
\end{equation}
\begin{equation}
  U^{\mbox{\small nnc}} = \sum^{\small\begin{minipage}{5em}\setlength{\baselineskip}{0mm} non-native\\contact\end{minipage}}_{j<i-3} 
  V^{\mbox{\small nnc}}_{ij}({\mathbf r}_{ij}),
\end{equation}
where z stands for $b$, $\theta$ or $\phi$,
and
$(1)$ and $(2)$ again indicate the reference conformations.
The vector ${\mathbf r}_{ij}={\mathbf r}_i-{\mathbf r}_j$ is the distance 
between the $i$th and $j$th C$_{\alpha}$,
where ${\mathbf r}_{i}$ is the position of the $i$th C$_{\alpha}$.
$b_{i}=|{\mathbf b}_{i}|=|{\mathbf r}_{i\,i+1}|$ 
is the virtual bond length between two adjacent C$_{\alpha}$.
$\theta_i$ is the angle between two adjacent virtual bonds, 
where $\cos\theta_i={\mathbf b}_{i-1}\cdot{\mathbf b}_{i}/b_{i-1}b_{i}$,
and $\phi_{i}$ is the $i$th dihedral angle around ${\mathbf b}_{i}$.
The first three terms of Eq. \ref{EQ:protein_potential} provide local interactions,
while the last two terms are interactions between non-local pairs that are distant along the chain.

\begin{figure}
\includegraphics{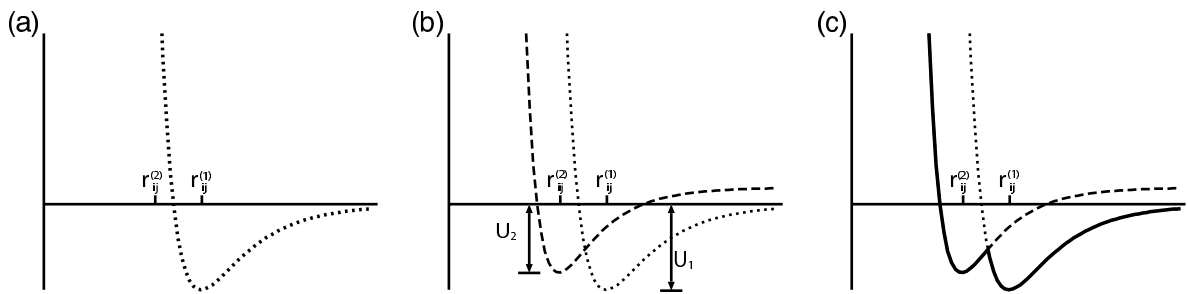}
 \caption{G\={o}-potential: The magenta line is dual G\={o}-potential energy profile 
 for the $ij$ pair. $C_{12}=U_2/U_1$.}
 \label{Fig:Go++Potential}
\end{figure}

For potential functions, $V_{\mbox{\small z}}^{(\alpha)}$, $V_{\mbox{\small nc}}^{(\alpha)}$, and $V_{\mbox{\small nnc}}^{(\alpha)}$, we use the same functions as Clementi {\it et al.} \cite{JMB-298_937}:
\begin{equation}
 V_{\mbox{\small b}}^{(\alpha)} (b_{i}) 
  = k_{\mbox{\small b}} (b_{i}-b_{i}^{(\alpha)})^2,
\end{equation}
\begin{equation}
 V_{\theta}^{(\alpha)} (\theta_i)
  = k_{\theta} (\theta_{i} -\theta_{i}^{(\alpha)})^2,
\end{equation}
\begin{eqnarray}
 V_{\phi}^{(\alpha)} (\phi_i)&=&k_{\phi} 
  \bigg[
   \left(1-\cos(\phi_{i} -\phi_{i}^{(\alpha)})\right)\nonumber\\
   &&+ \frac{1}{2} \left(1-\cos 3(\phi_{i} -\phi_{i}^{(\alpha)})\right)
 \bigg],
\end{eqnarray}
\begin{equation}
 V_{\mbox{\small nc}}^{(\alpha)} ({\mathbf r}_{ij})
 =  k_{\mbox{\small nc}}
  \left[
   5 \left(\frac{r^{(\alpha)}_{ij}}{r_{ij}}\right)^{12}
   - 6 \left(\frac{r^{(\alpha)}_{ij}}{r_{ij}}\right)^{10}
 \right],
\end{equation}
\begin{equation}
 V^{\mbox{\small nnc}}_{ij} ({\mathbf r}_{ij})
  =  k_{\mbox{\small nnc}}
  \left(\frac{C}{r_{ij}}\right)^{12},
\end{equation}
where the superscript $\alpha$ is 1 or 2 and represents the appropriate reference structure.
Parameters with superscript 1 or 2 are constants taken from the corresponding values 
in structure 1 or 2, respectively.
For the local interaction terms (bond length, bond angle, and dihedral angle), 
the potential energy for each set of beads takes the smaller of $V^{(1)}$ and $V^{(2)}$.
For example, the length of the $i$th bond is $b_{i}^{(1)}$ in structure 1 
and is $b_{i}^{(2)}$ in structure 2; therefore, the potential energy for this bond is 
$k_{\mbox{\small b}} \min\{(b_{i}-b_{i}^{(1)})^2,(b_{i}-b_{i}^{(2)})^2\}$.
We specify(define) that the $i$th and $j$th amino acids are in the ``native contact pair''
of structure 1 (or structure 2) if one of the non-hydrogen atoms 
in the $j$th amino acid is within 6.5 \AA
~of one of the non-hydrogen atoms in the $i$th amino acid at the structure 1 (or structure 2).
The interaction potential for each native-contact pair, $ij$,
takes the smaller of $V_{\mbox{\small nc}}^{(1)}({\mathbf r}_{ij})$ 
and $C_{12} V_{\mbox{\small nc}}^{(2)}({\mathbf r}_{ij})$.
$C_{12}$ is the ratio of the potential depth of structure 2 
to that of structure 1 (Fig.\ref{Fig:Go++Potential}).
Here, since we intend to perform simulations of the structural change
from structure 2 to structure 1, 
that is, we want to structure 1 as the final stable structure,
we assign a value smaller than unity to $C_{12}$ ($C_{12}=0.8$).
If a residue pair $ij$ is a native-contact pair in structure 1 but not in structure 2
(or vice versa), 
the interaction potential between the $i$th and $j$th residues is 
a Lennard-Jones potential, 
$V_{\mbox{\small nc}}^{(1({\mbox{\footnotesize or}}2))}({\mathbf r}_{ij})$,
with a single minimum.
Other relevant parameters are
$k_{\mbox{\small b}}=100.0$, 
$k_{\theta}=20.0$,
$k_{\phi}=1.0$, $k_{\mbox{\small nc}}=k_{\mbox{\small nnc}}=0.25$,
and $C=4.0$.
The cut-off length for calculating $V_{\mbox{\small nc}}^{(\alpha)}$ 
is taken to be $2r^{(\alpha)}_{ij}$.

To understand the mechanism of the actomyosin motor,
it is desirable to study the conformational change to the rigor state.
However, currently no X-ray crystal structure is currently available 
for a true rigor complex with actin;
therefore, we studied the structural relaxation
from the pre-power stroke state to the near-rigor state.
Structures 1 and 2 thus correspond to the near-rigor and  pre-power stroke structures, 
respectively.
We used 1Q5G \cite{NSB-10_826},
which is the nucleotide free structure of {\it Dictyostelium discoideum} myosin II,
as the near-rigor state.
Although a few nucleotide-free structures of myosin II have been determined,
only 1Q5G is regarded to be the near-rigor state, 
because both switch I and switch II are in the open position.
We chose 1VOM \cite{Biochem-35_5404} for the pre-power stroke structure.
It includes an ADP$\cdot$P$_{\mbox{\small i}}$ analog (ADP$\cdot$VO$_4$)
in the nucleotide binding site.
While 1Q5G consists of residues 2-765 without a gap region, 
1VOM includes only residues 2-747 and has gap regions 
where the structure has not been determined by X-ray crystallography.
Therefore we used the structures of residues 2-747 for simulations.
Potential functions of the near-rigor conformation
are assigned for local interactions in the gap region.

\begin{figure}
\begin{center}
\includegraphics{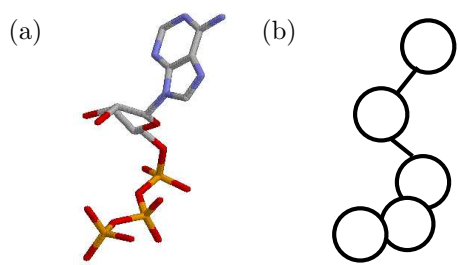}
\caption{(a) ATP and (b) coarse-grained ATP.}
 \label{Fig:ATP}
\end{center} 
\end{figure}

The nucleotide molecule (ADP$\cdot$VO$_4$ in 1VOM) is also included
in the simulation as a coarse-grained chain (Fig. \ref{Fig:ATP}).
The coarse-grained ADP$\cdot$VO$_4$ is represented as a short linear chain of five beads,
corresponding to a purine base, a sugar (ribose),
two phosphates, and VO$_4$ (a phosphate analog).
The intranucleotide interaction is defined as
\begin{eqnarray}
  U_{\mbox{\small n}} 
   &=&\!\! \sum_{k} k_{\mbox{\small b}}(|{\mathbf r}_{k+1}-{\mathbf r}_{k}|
   -|{\mathbf r}_{k+1}^{(2)}-{\mathbf r}_{k}^{(2)}|)^2\nonumber \\
   &&+ \sum_{k} k_{\mbox{\small b}}(|{\mathbf r}_{k+2}-{\mathbf r}_{k}|
   -|{\mathbf r}_{k+2}^{(2)}-{\mathbf r}_{k}^{(2)}|)^2.
\end{eqnarray}

The interaction potential between the protein and the nucleotide is similar 
to that between the non-local residues in the protein.
Here, we assume that only structure 2 includes the nucleotide:
therefore, the potential function has only a single well 
(the standard Lennard-Jones potential).
We specify that the $i$th residue of the protein and the $k$th bead in the nucleotide 
chain should be in ``native-contact'' in the pre-power stroke conformation
when one of the non-hydrogen atoms in the $k$th bead (base or sugar of P$_{\mbox{\small i}}$)
is within 4.5 \AA~ of one of the non-hydrogen atoms 
in the $i$th amino acid. 
The residues that form native-contacts with the nucleotide are called
nucleotide-contact residues:
\begin{eqnarray}
 U_{\mbox{\small p-n}}({\mathbf r}_{ik})
  &\!=\!& \sum^{\small\begin{minipage}{3em}\baselineskip=0ex native\\contact\end{minipage}}_{i,k}
 k_{\mbox{\small p-n}}
  \left[
   5 \left(\frac{r^{(2)}_{ik}}{r_{ik}}\right)^{12}
   \!\!-6 \left(\frac{r^{(2)}_{ik}}{r_{ik}}\right)^{10} \right]\nonumber\\
  && + \sum^{\small\begin{minipage}{4.7em}\setlength{\baselineskip}{0mm} non-native\\contact\end{minipage}}_{j<i-3} 
  k_{\mbox{\small nnc}}
   \left(\frac{C}{r_{ik}}\right)^{12},
\end{eqnarray}
where $i$ stands for the $i$th residue and $k$ stands for the $k$th bead in the nucleotide chain.

The dynamics of the proteins are simulated using the Langevin equation 
at a constant temperature $T$,
\begin{equation}
 m_{i}\dot{{\mathbf v}}_i = {\mathbf F}_i - \gamma {\mathbf v}_i + \xi_i
\end{equation}
where ${\mathbf v}$ is the velocity of the $i$th bead
and a dot represents the derivative with respect to time $t$ 
(thus, ${\mathbf v}_i=\dot{\mathbf r}_i$),
and ${\mathbf F}_i$ and $\xi_i$ are systematic and random forces 
acting on the $i$th bead, respectively.
The systematic force ${\mathbf F}_i$ is derived
from the effective potential energy $U$ and can be defined as 
${\mathbf F}_i=-\partial U/\partial {\mathbf r}_i$.
$\xi_i$ is a Gaussian white random force, which satisfies $\langle \xi_i \rangle = 0$ 
and $\langle \xi_i(t) \xi_j(t') \rangle=2\gamma T \delta_{ij} \delta(t-t') {\mathbf 1}$,
where the bracket denotes the ensemble average 
and ${\mathbf 1}$ is a $3 \times 3$ unit matrix.
We used an algorithm by Honeycutt and Thirumalai
\cite{Biopolymers-32_695}
for a numerical integration of the Langevin equation,
We used $\gamma=0.25$, $m_i$=1.0, and the finite time step $\Delta t=0.02$.

For a given protein conformation, $\Gamma$, 
we note that the native contact of structure 1 (or 2)
between $i$ and $j$ is formed
if the C$_{\alpha}$ distance $r_{ij}=|{\mathbf r}_{ij}|$ 
satisfies  $0.8 r^{(1)}_{ij}< r_{ij} < 1.2  r^{(1)}_{ij}$
($0.8 r^{(2)}_{ij}< r_{ij} < 1.2  r^{(2)}_{ij}$).

Simulations were started from the pre-power stroke structure.
The initial positions of residues in the gap regions of 1VOM were set randomly
under the condition that the bond length was 3.8 \AA.
The initial velocities of each bead was given to satisfy the Maxwell distribution.
The temperature was set lower than the folding temperature for structure 1.

We also ran a nucleotide-free and constrained simulation, 
in which the nucleotide was not explicitly included but 
the relative positions of the nucleotide-contact residues
were constrained by virtual bonds in all-to-all correspondence
to keep the pre-power stroke form.
The natural length of the virtual bonds $i$-$j$, $r_{ij}^{(2)}$,
is the C$_{\alpha}$ distance between the $i$th and $j$th residues 
in the pre-power stroke conformation (structure 2).
The total effective potential energy was 
$U_{\mbox{\small p}}+U_{\mbox{\small con}}$, and
\begin{equation}
 U_{\mbox{\small con}}=
 \sum^{\mbox{\small\begin{minipage}{3em}
		   \baselineskip=0ex nucl.-contact\end{minipage}}}_{j<i} 
		   k_{\mbox{\small con}}(r_{ij}-r_{ij}^{(2)})^2,
\end{equation}
where $k_{\mbox{\small con}}=1.0$.

\begin{acknowledgments}
We thank Shoji Takada and Toshio Yanagida for many helpful suggestions.
This work was supported 
by IT-program of the Ministry of Education, Culture, Sports, Science and Technology,
and Grant-in-Aid for Scientific Research (C) (17540383) 
from the Japan Society for the Promotion of Science.
\end{acknowledgments}

%\bibliography{motor.bib,f1-motor.bib,protein.bib,protein_nopdf.bib,phys.bib}

\end{article}

\end{document}